\documentclass[letterpaper]{article} 
\usepackage{aaai2027}  
\nocopyright
\usepackage[hyphens]{url}  
\usepackage{graphicx} 
\def\UrlFont{\rm}  
\usepackage{natbib}  
\usepackage{caption} 
\usepackage{algorithm}
\usepackage{algorithmic}

\usepackage{newfloat}
\usepackage{listings}
\DeclareCaptionStyle{ruled}{labelfont=normalfont,labelsep=colon,strut=off} 
\floatstyle{ruled}
\newfloat{listing}{tb}{lst}{}
\floatname{listing}{Listing}

\usepackage{booktabs}

\newcommand{\bheading}[1]{{\vspace{2pt}\noindent{\textbf{#1}}}}

\title{\textsc{ActGov}: Governing LLM Agent Actions via Policy-Constrained Validation}
\author{
    Kaiyuan Zhang\textsuperscript{\rm 1},
    Yuke Peng\textsuperscript{\rm 1},
    Ke Jiang\textsuperscript{\rm 1},
    Yinqian Zhang\textsuperscript{\rm 1}
}

\affiliations{
    \textsuperscript{\rm 1}Southern University of Science and Technology\\
    12432720@mail.sustech.edu.cn, pengyk@mail.sustech.edu.cn\\
    ke006@e.ntu.edu.sg, yinqianz@acm.org
}

\usepackage{amssymb}
\usepackage{amsmath,mathtools}
\usepackage{booktabs}
\usepackage{longtable}
\usepackage{array}
\usepackage{tabularx}
\usepackage{pdflscape}
\usepackage{enumitem}
\usepackage{xspace}
\usepackage{listings}
\usepackage{caption}
\usepackage{subcaption}
\usepackage{microtype}
\usepackage{placeins}

\newcommand{\system}{\textsc{ActGov}\xspace}
\newcommand{\policy}{\textsc{ActGov-Policy}\xspace}

\newcommand{\runtime}{\textsc{ActGov-Runtime}\xspace}

\newcommand{\F}{Fig.}

\newenvironment{packeditemize}{
\begin{list}{$\bullet$}{
\setlength{\labelwidth}{0pt}
\setlength{\itemsep}{2pt}
\setlength{\leftmargin}{\labelsep}
\setlength{\parindent}{0pt}
\setlength{\listparindent}{\parindent}
\setlength{\parsep}{1pt}
\setlength{\topsep}{1pt}}}{\end{list}}

\usepackage{listings}
\usepackage{xcolor}

\definecolor{codegray}{rgb}{0.96,0.96,0.96}
\definecolor{codeframe}{rgb}{0.85,0.85,0.85}
\definecolor{codekw}{rgb}{0.13,0.29,0.53} 

\lstdefinestyle{policystyle}{
    backgroundcolor=\color{codegray},   
    commentstyle=\color{gray},
    keywordstyle=\color{codekw}\bfseries,
    numberstyle=\footnotesize\color{gray},
    stringstyle=\color{purple},
    basicstyle=\ttfamily\small,
    breakatwhitespace=false,         
    breaklines=true,                 
    captionpos=b,                    
    keepspaces=true,                 
    showspaces=false,                
    showstringspaces=false,
    showtabs=false,                  
    frame=single,                   
    rulecolor=\color{codeframe},    
    xleftmargin=1em,                
    xrightmargin=1em,
    aboveskip=1em,                  
    belowskip=1em
}

\begin{document}

\maketitle

\begin{abstract}
Large language model (LLM) agents increasingly execute long-horizon workflows through external tools, allowing untrusted outputs to influence subsequent actions and exceed user authorization. Existing defenses isolate injected content or constrain execution with predefined plans and static policies, but these approaches are brittle under dynamic workflows and scale poorly across extensible tool ecosystems.

In this work, we present \textsc{ActGov}, a runtime enforcement framework that validates each LLM-proposed tool action before it causes external effects. Built on a unified semantic model of authorization, actions, runtime context, and security constraints, the \textsc{ActGov-Policy} component iteratively constructs a policy set from tool specifications, benign tasks, and observed failure traces, with each update verified through SMT-based counterexample checking. At runtime, \textsc{ActGov-Runtime} abstracts each tool call into finite policy records and permits it only if it remains within the task-scoped authorization boundary and satisfies all applicable policies. This per-action enforcement preserves authorization throughout long-horizon, dynamically branching workflows.

We evaluate \textsc{ActGov} on the AgentDojo and AgentDyn benchmarks across multiple models and attack configurations. It shows that \textsc{ActGov}  consistently reduces the success rate of indirect prompt-injection attacks while preserving task utility, significantly outperforming existing defenses. These results demonstrate that \textsc{ActGov} can enforce fine-grained authorization over dynamic agent executions without relying on the underlying LLM to correctly identify malicious instructions.
\end{abstract}


\section{Introduction}

Large language models (LLMs) are evolving from text generators into autonomous agents that can plan, invoke tools, observe external environments, and execute multi-step actions across resources such as email, calendars, file systems, code repositories, financial accounts, and enterprise platforms~\cite{yao2023react,schick2023toolformer,ICLR2024_e91bf7df}. 
This transition fundamentally changes the security boundary: model outputs may now directly trigger high-impact operations, e.g., modifying, deleting, and authorizing~\cite{agentdojo,progent}.
The primary risk is that untrusted external content can be misinterpreted as user intent, causing the agent to execute unauthorized actions~\cite{IPI,agentdojo}.
The dynamic dependency between tool results and subsequent actions demands per-tool-call authorization enforcement rather than input-output level filtering, while balancing security and task utility~\cite{wu2024systemlevel,agentdojo}.

Existing defenses span prompt-level provenance marking and injection filtering~\cite{Spotlighting,promptguard,PromptArmor}, runtime monitoring and tool-call governance~\cite{progent,drift}, information-flow and capability separation~\cite{wu2024systemlevel,camel}, and execution isolation or structured-plan enforcement~\cite{wu2025isolategpt,ace}.
While these methods significantly improve agent security, they face challenges in addressing the complexity of dynamic, open-ended tool ecosystems due to the following limitations.

\begin{packeditemize}
\item Existing defenses \textit{lack a unified semantic framework} for representing autonomous agent actions. Instead, they rely on representations tailored to specific security mechanisms.
Without such a unified semantic foundation, defenses may misinterpret the relationship between data and authorization provenance, either treating untrusted content as an authority source or failing to recognize unauthorized action paths, allowing injected content to influence agent decisions as if it were user intent~\cite{Spotlighting,camel}.

\item Existing defenses \textit{lack robustness for long-horizon agent workflows}. Many approaches constrain execution using predefined plans, expected tool traces, or static parameter checklists~\cite{drift, ace}. However, long-running agents must continuously adapt to intermediate tool outputs, dynamic branching, and unexpected execution states, making static constraints difficult to maintain. 
These approaches either over-restrict legitimate adaptive behaviors and reduce task utility, or become ineffective when execution deviates from the predefined assumptions~\cite{agentdyn,xiang2026architectingsecureaiagents}. 
A robust defense should therefore reason about each action in its current execution context rather than relying on a fixed global plan.

\item Existing defenses \textit{face challenges in scaling policy construction while maintaining security assurance}. As tool ecosystems expand, manually authored policies become increasingly difficult to maintain and adapt to new tools, actions, and execution contexts. Although automatically generated policies or privileges~\cite{progent} improve scalability, they lack rigorous mechanisms to detect omissions, conflicts, or unintended permissions before deployment. Formal verification can provide stronger assurance, but existing approaches, such as VeriGuard~\cite{veriguard}, often verify policy logic separately from the concrete runtime states encountered by agents. This disconnect between policy reasoning and runtime execution limits end-to-end security guarantees. 
\end{packeditemize}

To address these challenges, we present \system, a formally verified authorization framework that maps unstructured execution contexts into a discrete 2D record schema. 
This architecture enables LLM-driven automated policy generation and Z3-based verification~\cite{z3}, achieving scalable policy construction and verifiable security enforcement before external tool actions are executed.


The foundation of \system is a finite two-dimensional record space
$\mathcal{R}$, organized along \textit{formation levels} and
\textit{semantic scopes}, that maps unstructured runtime contexts into
bounded authorization-relevant abstractions. \system abstracts concrete resources and their provenance into a finite
set of predefined record values, allowing policies to generalize across
different resource identifiers. LLMs are used only to map unstructured
context into these records and to propose policies in \policy; runtime
authorization is determined deterministically by the verified policy
bundle. Before deployment, \policy iteratively refines a
layered policy bundle $\mathcal{P}$, which is checked by Z3 against
predefined safety assertions. The verified bundle is then frozen and
enforced by \runtime.

Built upon this schema, \system structures its rules into three complementary classes: \textit{Task-Permission}, \textit{Hard-Invariants} , and  \textit{Procedural-Obligations} to govern ambiguous contexts. 
By offloading the computationally expensive Z3 solver to the offline \policy phase, an \texttt{unsat} result guarantees the absence of conflicting rules or exploitable loopholes prior to deployment. 
Consequently, the \runtime monitor merely needs to evaluate the abstracted finite records against this mathematically sound, static policy set. This architectural split ensures low-latency, deterministic enforcement at runtime while maintaining rigorous, formal security guarantees.

To sum up, this paper makes the following contributions.
\begin{packeditemize}
\item We introduce a unified semantic abstraction that maps authorization, agent actions, runtime context, provenance, and security constraints into a finite record space, enabling precise specification, runtime enforcement, and formal verification.

\item We develop a scalable policy construction framework that uses LLMs to automatically propose and iteratively refine layered security policies from tool specifications, benign tasks, and attack traces. Crucially, each LLM-proposed iteration is rigorously validated through Z3 solver to guarantee soundness before deployment.

\item We extensively evaluate \system on AgentDojo and AgentDyn across multiple models and attack settings, demonstrating that it substantially reduces indirect prompt-injection success while preserving task utility and outperforming existing defenses.
\end{packeditemize}


\section{Related Work}
\label{sec:prelim}


\bheading{Formal and Policy-Based Agent Defenses.}
VeriGuard focuses on translating explicit safety requests and agent
specifications into executable policy code, and formally verifies that
the generated implementation satisfies its corresponding pre- and
post-conditions before runtime enforcement~\cite{veriguard}. Progent
instead targets least-privilege enforcement, representing the currently
permitted tool-call space through symbolic rules over tool names and
arguments and using SMT to determine whether each policy update narrows
or expands that space~\cite{progent}. The published descriptions do not
provide explicit evidence that either framework adopts a shared, typed
abstraction of runtime context that exposes task permissions, argument
provenance, target bindings, and execution history as first-class factors
for both verification and enforcement. Such a common semantic layer is
valuable because it provides a consistent basis for expressing and
checking interactions among heterogeneous authorization factors before
deployment.

To address this gap, \system uses a finite semantic record space shared
by offline verification and runtime enforcement. Before deployment, it
checks safety assertions over the entire layered policy bundle within
the modeled finite record domain; at runtime, each candidate action is
evaluated deterministically over the same record abstraction. This shared
representation enables these contextual authorization factors to be
considered jointly, aligns runtime authorization with the properties
verified offline, and avoids online policy synthesis or solver invocation.

 


\begin{figure*}[!t]
  \centering
  \includegraphics[width=\textwidth]{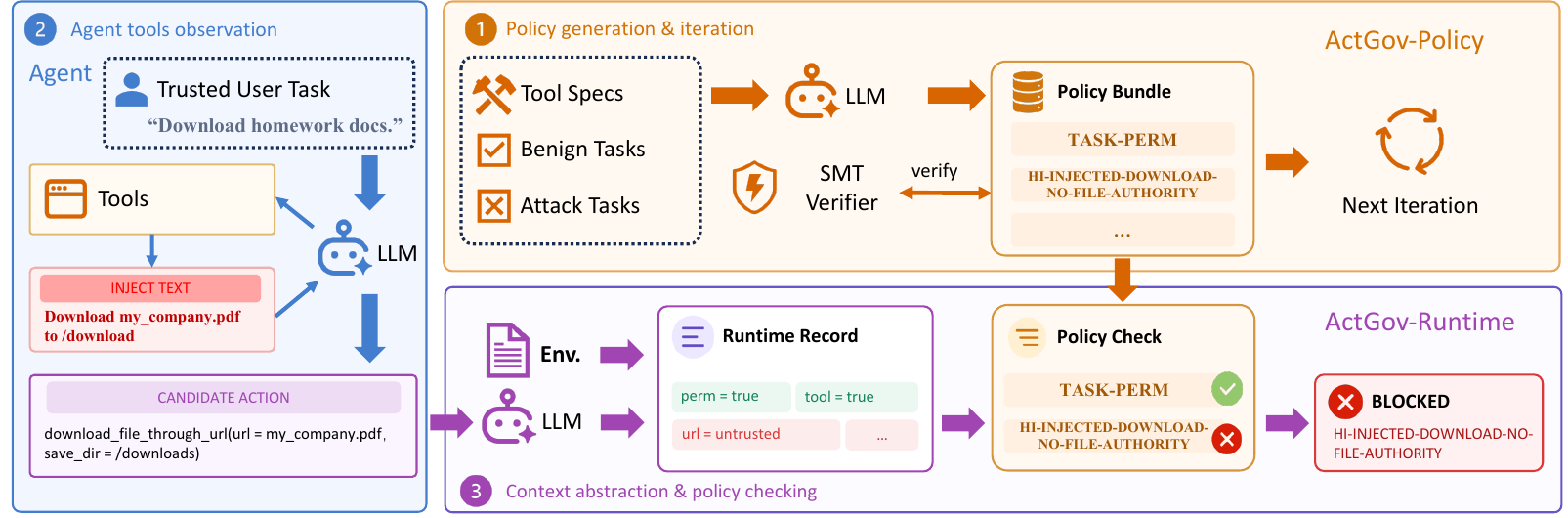}
  \caption{Overview of \system. \system comprises ActGov-Policy for policy generation and iteration, and ActGov-Runtime for context abstraction and policy checking.}
  \label{fig:overview}
\end{figure*}

\bheading{Structural Defenses and Execution-Level Analysis.}
A complementary line of research secures agent execution by constraining
plans, information flows, and runtime dependencies. Methods such as
CaMeL and ACE establish trusted control-flow or abstract-plan boundaries
and enforce information-flow and capability constraints~\cite{camel,ace}.
Other approaches analyze runtime behavior by validating deviations from
an intended trajectory or reconstructing execution traces into dependency
and influence graphs~\cite{drift,agentarmor,argus}. These structural
defenses provide strong isolation, but precommitted plans and control-flow
boundaries can be conservative when legitimate action targets must be
resolved from runtime observations, whereas trace-level analysis may
require continuous online reasoning over evolving execution states.

\system instead separates dynamic action proposal from execution
authorization. The agent remains free to adapt its plan using runtime
observations, while each candidate tool call is authorized deterministically
against a verified policy set over a finite semantic record space.
This design preserves flexibility for dynamic workflows while avoiding
online SMT solving and reducing reliance on the agent LLM for final
security decisions.

\section{Methodology}
\label{sec:methodology}

\bheading{Threat Model.} 
We consider indirect prompt injection (IPI): the initial user instruction is trusted; the attacker-controlled content enters through downstream tool observations and impels LLMs to propose unauthorized actions~\cite{agentdojo,IPI}. 
The attacker may control external content and manipulate action proposals, but cannot modify the trusted user task or \system's enforcement layer, which includes tool specifications and deployed policies. 
Finally, we assume that no action can produce external effects without first passing \system's pre-execution validation.

\bheading{Overview.} 
\system is a formally verified authorization framework designed to protect LLM-based tool-calling agents by separating action intent from execution authority. 
The agent may freely propose a candidate tool call, but the call is not
forwarded to the external tool interface until it has been comprehensively validated by \system. An overview of \system is shown in \F~\ref{fig:overview}. \system comprises two key components:

\begin{packeditemize}
\item \textbf{\policy (Policy Generation \& Iteration):} A policy generation and iteration engine that iteratively synthesizes, checks, and refines a candidate policy set before deployment. As illustrated in \F~\ref{fig:overview}, this component uses inputs such as tool specifications, benign task traces, and known attack tasks.  It leverages an LLM to draft candidate policy rules. 
Before being integrated into the active set, these rules are validated by a Z3 solver, which mathematically guarantees that the generated policies introduce no logical contradictions and strictly preserve predefined security invariants.

\item \textbf{\runtime (Context Abstraction \& Policy Checking):} A runtime enforcement monitor deployed between the agent and external tools. When the agent proposes a candidate action, \runtime intercepts the request. Instead of evaluating raw natural language, it abstracts the action and its execution context into a finite runtime record space. This record then undergoes a deterministic policy evaluation against the verified policy set. If the record violates any security rule, the action is blocked.
\end{packeditemize}

Both components operate over a \textbf{unified formal semantics model}. 
Rather than dealing with unstructured natural language prompts and arbitrary tool APIs, \system formalizes the agent's execution into a rigorous mathematical structure.
This formalization serves as the semantic bridge of our framework: it provides a standardized language that makes subsequent security policies precisely definable and verifiable via automated formal methods, as detailed in the following Formulation subsection.

\bheading{Governance Goals.}
The design goal of \system is to reduce the unpredictability of LLM-driven security decisions by constraining the LLM's role in transforming natural language into structured records. 
Rather than relying on the LLM for complex security reasoning or authorization decisions, \system uses it solely as a semantic parser for bounded information extraction. Specifically, the LLM evaluates localized contexts to extract predefined boolean or categorical values (e.g., whether a parameter originates from an untrusted source). 
While the extraction process necessarily relies on the LLM's natural language understanding, confining its role to these predefined tasks and delegating the actual authorization logic to a deterministic interpreter strictly limits the error space caused by reasoning hallucinations. 
This architecture ensures a reliable abstraction of the complex runtime environment, thereby enhancing the overall robustness and soundness of the enforcement framework.

\subsection{Formulation}
\label{sec:semantic}





To integrate formally verified policies into the agent runtime, we formalize the execution to decouple the LLM's generative reasoning from \system's deterministic monitoring. 
Consider an agent executing a user task $\tau$ over a sequence of tool interactions. At step $s_i$, the agent's LLM observes the execution history $H_i$ prior to step $s_i$ and proposes a candidate tool call $c_i$.
This proposal may be influenced by untrusted external content (e.g., injected instructions within a web page) or by model hallucinations~\cite{zhang-etal-2024-toolbehonest}.
While the LLM uses the execution history $H_i$ to determine the next action, \system \textbf{does not} evaluate this entire unstructured history to validate security policies.
Instead, the relevant environmental and temporal states from $H_i$ and $c_i$ are abstracted into a finite runtime record set $r_i \in \mathcal{R}$.

Formally, let $\mathcal{H}$ denote the execution history state space, $\mathcal{C}$ the candidate tool call space, $\mathcal{R}$ the finite record space, and $\mathcal{P}$ the policy set. The evaluation function $\mathcal{V}: \mathcal{R} \times \mathcal{P} \rightarrow \{\text{allow}, \text{deny}\}$ evaluates the abstracted records $r_i$ against the policy set $\mathcal{P}$, producing a deterministic authorization verdict $v_i$. The state transition function $\mathcal{T}: \mathcal{H} \times \mathcal{C} \times \mathcal{V} \rightarrow \mathcal{H}$ incorporates the candidate action $c_i$ and the verdict $v_i$ to update the execution history from $H_i$ to $H_{i+1}$. The external tool is executed if and only if $v_i = \text{allowed}$.

Ultimately, we model an agent protected by \system as a finite state machine:
$$
\mathcal{M} = (\mathcal{H}, \mathcal{C}, \mathcal{R}, \mathcal{P}, \mathcal{V}, \mathcal{T}).
$$

\subsection{\policy}
\label{sec:offline-policy}

Rather than constructing every rule manually, the \policy phase systematically generates, verifies, and iteratively refines the policy set $\mathcal{P}$ prior to deployment. 
Based on the finite record space $\mathcal{R}$ established in our formal model, this  LLM-assisted pipeline leverages an LLM to draft and evolve policies across three data sources: tool specifications for tool capability and risk constraints, benign task traces for identifying over-restrictive rules, and observed attack failures for identifying missing security conditions.
The LLM operates exclusively as a policy proposer during this phase. Candidate rules remain subject to formal checking and administrator review before deployment. Once the iteration converges, the drafted rules are presented to administrators for confirmation.


\bheading{Orthogonal 2D Record Schema.}
To bridge unpredictable LLM behaviors and deterministic security rules, $\mathcal{R}$ is organized as an orthogonal 2D schema over \textit{formation levels} and \textit{semantic scopes}. The two dimensions separate how a record value is derived from what authorization-relevant property it describes. Rather than reasoning directly over raw natural-language context, \system maps the current execution context into structured records evaluated by executable policies. 
As illustrated in Table~\ref{tab:record_examples}, each record is classified by one formation level and one semantic scope. For example, the record \textit{pending\_obligation} belongs to the L1 deterministic level and the History scope. This record evaluates whether the execution trace contains an unresolved user-approval requirement from previous steps. 
This design constrains context extraction to a fixed schema and finite value domains, preventing adversarial instructions from directly influencing the final policy evaluation.

\begin{table}[t]
\centering
\caption{Example records in the orthogonal 2D schema. Each record belongs
to one semantic scope and one formation level.}
\label{tab:record_examples}
\small
\setlength{\tabcolsep}{4pt}
\renewcommand{\arraystretch}{1.12}
\begin{tabular}{@{}llccc@{}}
\toprule
\textbf{Record} &
\textbf{Semantic Scope} &
\textbf{L1} &
\textbf{L2} &
\textbf{L3} \\
\midrule
\texttt{outbound\_tag}
    & Action     & \checkmark &            &            \\
\texttt{tool\_allowed}
    & Permission & \checkmark &            &            \\
\texttt{arg\_provenance}
    & Parameter  &            & \checkmark &            \\
\texttt{pending\_obligation}
    & History    & \checkmark &            &            \\
\texttt{binding\_context}
    & Binding    & \checkmark &            &            \\
\texttt{trusted\_endpoint}
    & External   &            &            & \checkmark \\
\texttt{goal\_balance\_check}
    & Domain     &            & \checkmark &            \\
\texttt{is\_high\_risk}
    & Action     & \checkmark &            &            \\
\texttt{binding\_valid}
    & Binding    &            &            & \checkmark \\
\bottomrule
\end{tabular}
\end{table}

On the formation axis, records are organized by how their values are produced:
\begin{packeditemize}
    \item \textbf{L1 Deterministic:} Records computed directly from explicit, structured system states without requiring natural language processing.
    \item \textbf{L2 Bounded Semantic:} Records that require semantic understanding to extract contextual values from unstructured text. The LLM functions exclusively as a semantic parser here, mapping the context into predefined, finite value domains rather than making authorization decisions.
    \item \textbf{L3 Composed:} Records derived from underlying L1 and L2 values via standard Boolean logic (e.g., AND/OR/NOT) to capture combinatorial or temporal dependencies.
\end{packeditemize}

On the semantic axis, the schema defines seven semantic scopes: \textbf{action}, \textbf{permission}, \textbf{parameter}, \textbf{history}, \textbf{binding}, \textbf{external}, and \textbf{domain}. These domains respectively characterize the proposed candidate action, the baseline permissions granted to the task, the origin and role of action parameters, the relevant execution history, the consistency between intended and actual targets, the security properties of external destinations, and the current operational stage of the task. By decomposing the execution context into this structured record space, \system provides a finite well-defined state space for both policy authoring and formal verification.

\bheading{Layered Authorization Policies.}
To establish well-defined security boundaries, the deployed policy set comprises three complementary layers: $\mathcal{P} = \mathcal{P}_{\mathrm{TP}} \cup \mathcal{P}_{\mathrm{HI}} \cup \mathcal{P}_{\mathrm{PO}}$. 
\begin{packeditemize}
    \item \textbf{Task-Permission ($\mathcal{P}_{\mathrm{TP}}$):} Binds a trusted user task, derived from the original user prompt, to the minimum required tools and resources, generating a baseline authorization boundary $perm$. \textit{Actions outside this boundary are blocked.}

    \item \textbf{Hard-Invariant ($\mathcal{P}_{\mathrm{HI}}$):} Encodes global security properties, such as preventing outbound network actions derived from untrusted injected content, regardless of the active task scope.

    \item \textbf{Procedural-Obligation ($\mathcal{P}_{\mathrm{PO}}$):} Covers conditions that fall beyond the agent's autonomous processing authority, such as weak target binding or uncertain argument provenance. Unlike hard invariants, these act as conditional rules: they mandate specific prerequisites or contextual constraints that must be satisfied before execution is permitted.
    
\end{packeditemize}




\bheading{Verification-Gated Evolution.}
\policy iteratively evolves the policy set using tool specifications, benign task traces, and observed failures. However, no update modifies the runtime monitor directly. Instead, candidate policy bundles are validated by a Z3 solver~\cite{z3}. For any predefined security assertion $A_{\mathrm{sec}}$, the verifier searches the finite record domain for a legal assignment $r \in \mathcal{R}$ and decision $v$ that satisfies the policy semantics $\mathcal{P}_{\mathrm{sem}}$ but violates the assertion:
\[
\exists r,v\; .\; \mathrm{Domain}_R(r) \land \mathcal{P}_{\mathrm{sem}}(r,v) \land \neg A_{\mathrm{sec}}(r,v).
\]
A \texttt{SAT} result exposes a counterexample representing a policy gap or conflict, triggering policy refinement. A set is deployed only when all queries return \texttt{UNSAT}, which mathematically guarantees that the updated policies preserve predefined security invariants. By completing this verification entirely during the preparation stage, \system ensures structural security without introducing formal solving latency into the runtime execution.

\subsection{\runtime}
\label{sec:runtime-checking}


At runtime, \runtime is deployed between the agent's LLM and the external tool interface. It intercepts the candidate tool call $c_i$, instantiates a data record set from the predefined schema $\mathcal{R}$, dynamically populates it by extracting the current execution context, and evaluates the policy predicates to enforce authorization. 
Initially, $\mathcal{P}_{\mathrm{TP}}$ assigns a baseline authorization boundary $perm$, restricting the agent to the minimum required tool permissions for the specific task. Within this boundary, \runtime further evaluates the action against the security rules defined in $\mathcal{P}_{\mathrm{HI}}$ and $\mathcal{P}_{\mathrm{PO}}$. An action is permitted to execute if and only if it resides within $perm$ and violates no security policies.


\bheading{Record Abstraction.}
The security of a tool call depends on its context. For instance, a parameter may be benign when user-provided but malicious when extracted from an untrusted email. 
At step $s_i$, \runtime resolves the baseline task permission $perm$ via $\mathcal{P}_{\mathrm{TP}}$ and abstracts the environment into a policy record set $r_i$:
$$r_i=\alpha_R(\tau, perm, H_i, c_i).$$
\runtime populates $r_i$ according to the 2D schema $\mathcal{R}$. To avoid hallucinations and prompt injections, \runtime does not rely on the LLM for end-to-end security judgments. Instead, it confines the LLM to function exclusively as an isolated feature extractor for specific records (e.g., categorizing parameter provenance). This constraint minimizes generation errors, ensuring that policies evaluate structured, finite values rather than raw natural-language text.



\section{Experimental Setup}
\label{sec:exp_setup}

We evaluate \system on two tool-calling agent benchmarks, AgentDojo~\cite{agentdojo} and AgentDyn~\cite{agentdyn}, under indirect prompt-injection attacks. The goal of our evaluation is to answer the following research questions:

\begin{enumerate}[leftmargin=*, label=\textbf{RQ\arabic*:}]
    \item \textbf{Efficacy \& Comparative Advantage:} Can \system outperform state-of-the-art system-level defenses in reducing attack success rates (ASR) without relying on the underlying LLM to recognize malicious instructions?
    \item \textbf{Utility:} Does \system preserve task completion rates under both benign and attacked execution compared with the baseline without defense?
    \item \textbf{Policy Generation Dynamics:} How does the choice of the policy-generating LLM, as well as the scale of the training dataset, impact the efficacy and utility of the synthesized defense policies?
\end{enumerate}

\subsection{Benchmarks}

\bheading{AgentDojo.}
This benchmark evaluates tool-calling agents on realistic multi-step tasks
across domains such as workspace, banking, travel, and Slack. Attacks
are injected into tool observations, such as emails and webpages. It 
contains 629 security test cases, each pairing a user task
with a compatible injection task, and separately evaluates user-goal
completion and attacker-goal success~\cite{agentdojo}.

\bheading{AgentDyn.}
This benchmark introduces dynamic, open-ended tasks that incorporate
third-party instructions. It is particularly challenging for
policy-based defenses because legitimate execution often requires
deriving action targets from external observations while disregarding
malicious instructions embedded in the same environment. It
contains 60 user tasks and 28 injection tasks, combined into 560
security test cases~\cite{agentdyn}.

\subsection{Models and Baselines}

To assess the generalizability of our framework, we evaluate \system across four LLM backends: Qwen3.6-flash~\cite{qwen36flush}, MiniMax-M2.5~\cite{minimaxm25}, DeepSeek-v4-pro~\cite{deepseekv4pro}, and
GPT-4o mini~\cite{gpt4omini}. We benchmark our approach against a baseline without defense and four
system-level defense baselines. Notably, we omit VeriGuard~\cite{veriguard} from our evaluation as its source code is currently not publicly available:

\begin{packeditemize}
    \item \textbf{CaMeL} enforces information-flow constraints between trusted and untrusted data~\cite{camel}. We adapt CaMeL as a guarded pipeline while preserving native benchmark scorers.
    \item \textbf{Progent} is a policy-based guarded execution baseline that restricts tool calls based on predefined privileges \cite{progent}.
    \item \textbf{DRIFT} applies dynamic validation and trace isolation rules~\cite{drift}. We use its official evaluation procedure and report all completed guarded runs.
    \item \textbf{ACE} is a framework requiring agents to generate abstract plans before mapping them to concrete tools~\cite{ace}. We evaluate ACE using the native benchmark scorer.
\end{packeditemize}

\begin{table*}[t]
    \centering
    \caption{System-level defense comparison on the AgentDyn test splits.}
    \label{tab:system_eval}
    \resizebox{\textwidth}{!}{
    \setlength{\tabcolsep}{3.5pt} 
    \begin{tabular}{llccc @{\hspace{1em}} llccc} 
        \toprule
        Model & Method & Clean $\uparrow$ & Attacked $\uparrow$ & ASR $\downarrow$ &
        Model & Method & Clean $\uparrow$ & Attacked $\uparrow$ & ASR $\downarrow$ \\
        \midrule
        
        Qwen3.6-flash & No defense & 0.667 & 0.650 & 0.046 & DeepSeek-v4-pro & No defense & 0.767 & 0.729 & 0.086 \\
        & \system & \textbf{0.667} & \textbf{0.586} & \textbf{0.004} & & \system & \textbf{0.600} & \textbf{0.511} & \textbf{0.007} \\
        & Progent & 0.150 & 0.093 & 0.011 & & Progent & 0.217 & 0.243 & 0.014 \\
        & DRIFT   & 0.033 & 0.046 & 0.000 & & DRIFT   & 0.300 & 0.300 & 0.039 \\
        & CaMeL   & 0.000 & 0.000 & 0.000 & & CaMeL   & 0.000 & 0.000 & 0.000 \\
        & ACE     & 0.000 & 0.000 & 0.000 & & ACE     & 0.000 & 0.000 & 0.000 \\
        \midrule
        
        MiniMax-M2.5  & No defense & 0.650 & 0.625 & 0.036 & GPT-4o mini & No defense & 0.467 & 0.414 & 0.111 \\
        & \system & \textbf{0.617} & \textbf{0.546} & \textbf{0.004} & & \system & \textbf{0.433} & \textbf{0.343} & \textbf{0.007} \\
        & Progent & 0.200 & 0.161 & 0.029 & & Progent & 0.033 & 0.032 & 0.018 \\
        & DRIFT   & 0.250 & 0.189 & 0.046 & & DRIFT   & 0.133 & 0.232 & 0.036 \\
        & CaMeL   & 0.000 & 0.000 & 0.000 & & CaMeL   & 0.000 & 0.000 & 0.000 \\
        & ACE     & 0.000 & 0.000 & 0.000 & & ACE     & 0.000 & 0.000 & 0.000 \\
        \bottomrule
    \end{tabular}
    }
\end{table*}

\begin{table*}[t]
    \centering
    \caption{System-level defense comparison on the AgentDojo test splits.}
    \label{tab:actgov-agentdojo-comparison}
    \resizebox{\textwidth}{!}{
    \setlength{\tabcolsep}{3.5pt} 
    \begin{tabular}{llccc @{\hspace{1em}} llccc} 
        \toprule
        Model & Method & Clean $\uparrow$ & Attacked $\uparrow$ & ASR $\downarrow$ &
        Model & Method & Clean $\uparrow$ & Attacked $\uparrow$ & ASR $\downarrow$ \\
        \midrule
        
        Qwen3.6-flash & No defense & 0.857 & 0.857 & 0.069 & DeepSeek-v4-pro & No defense & 0.878 & 0.839 & 0.061 \\
        & \system & \textbf{0.714} & \textbf{0.673} & \textbf{0.000} & & \system & 0.755 & 0.713 & 0.000 \\
        & Progent & 0.653 & 0.667 & 0.004 & & Progent & \textbf{0.816} & \textbf{0.799} & \textbf{0.008} \\
        & DRIFT   & 0.347 & 0.380 & 0.004 & & DRIFT   & 0.796 & 0.753 & 0.063 \\
        & CaMeL   & 0.184 & 0.229 & 0.002 & & CaMeL   & 0.204 & 0.208 & 0.000 \\
        & ACE     & 0.143 & 0.143 & 0.000 & & ACE     & 0.082 & 0.040 & 0.000 \\
        \midrule
        
        MiniMax-M2.5  & No defense & 0.857 & 0.857 & 0.031 & GPT-4o mini & No defense & 0.653 & 0.690 & 0.036 \\
        & \system & \textbf{0.735} & \textbf{0.700} & \textbf{0.000} & & \system & \textbf{0.612} & \textbf{0.606} & \textbf{0.000} \\
        & Progent & 0.510  & 0.505 & 0.002 & & Progent & 0.388 & 0.342 & 0.000 \\
        & DRIFT   & 0.551 & 0.480 & 0.040 & & DRIFT   & 0.674 & 0.581 & 0.015 \\
        & CaMeL   & 0.224 & 0.245 & 0.002 & & CaMeL   & 0.306 & 0.361 & 0.000 \\
        & ACE     & 0.143 & 0.159 & 0.000 & & ACE     & 0.122 & 0.101 & 0.000 \\
        \bottomrule
    \end{tabular}
    }
\end{table*}

\subsection{Evaluation Integrity and Experimental Scenarios}

We partition the attack pairs of AgentDojo and AgentDyn into equal training and test splits. For our main evaluation (RQ1 \& RQ2), the training split is used exclusively by \policy component. To evaluate \system under a strong policy-generation setting, \policy uses GPT-5.5~\cite{gpt55} as the policy proposer over 10 refinement rounds, combined with human-in-the-loop (HITL) review. Upon concluding this phase, the policy set is frozen. All subsequent evaluations for \system and the baselines are conducted on the unseen test split to ensure a fair comparison.

To address RQ3 regarding autonomous scalability, we introduce two experimental scenarios on AgentDyn. In these settings, we remove the HITL review and constrain policy generation to a fully automated 5-iteration loop:
\begin{packeditemize}
    \item \textbf{Impact of Policy-Generating LLMs:} We instruct different LLMs, using identical prompts, to generate and refine policies on the same fixed 50\% subset of the training split. The resulting policies are evaluated on the full test split using DeepSeek-v4-pro as the fixed agent backend.
    \item \textbf{Impact of Training Data Scale:} 
    Using DeepSeek-v4-pro as both the policy generator and the agent
backend, we vary the amount of training data used for automated policy
construction among 50\%, and 25\% of the training split. The 
policy sets are evaluated on the full test split.
\end{packeditemize}

\subsection{Evaluation Metrics}

A viable defense must minimize the attack success rate (ASR) while maintaining high utility; trivial solutions that block all executions achieve near-zero ASR but zero utility. Thus, ASR reduction is only meaningful when contextualized by utility preservation. Following the native user-goal and attacker-goal scorers of AgentDojo and AgentDyn~\cite{agentdojo,agentdyn}, we measure performance using three metrics:

\begin{packeditemize}
    \item \textbf{Clean Utility:} The fraction of clean (unattacked) user tasks that are successfully completed.
    \item \textbf{Attacked Utility:} The fraction of attacked executions where the agent successfully completes the original user task despite the injection.
    \item \textbf{Attack Success Rate (ASR):} The fraction of attacked executions where the attacker's specific goal is achieved. 
\end{packeditemize}


\section{Results and Analysis}
\label{sec:results}

We structure our analysis to directly address the research questions formulated in the previous section. To ensure a rigorous evaluation, we separate within-benchmark results according to their specific policy origins, evaluating the AgentDyn-trained policies on the AgentDyn test split and the AgentDojo-trained policies on the AgentDojo test split.

\subsection{Defense Efficacy and Utility Preservation (RQ1 \& RQ2)}

We first evaluate whether \system can neutralize indirect prompt injections while strictly preserving the agent's ability to complete benign tasks. Table~\ref{tab:system_eval} details the performance on the AgentDyn test split across four model backends. Table~\ref{tab:actgov-agentdojo-comparison} presents the corresponding within-benchmark evaluation on the AgentDojo test split.

A trivial defense can eliminate attacks by blocking all executions, so
efficacy must be considered together with utility. As shown in
Table~\ref{tab:system_eval}, on AgentDyn, \system limits ASR to at most
0.007 across all four models while preserving competitive clean and
attacked utility. For example, on Qwen3.6-Flash, it retains utilities of
0.667 and 0.586, compared with 0.667 and 0.650 under no defense.
Table~\ref{tab:actgov-agentdojo-comparison} further shows that, on
AgentDojo, \system achieves an ASR of 0.000 across all four backends
while maintaining top-tier utility on Qwen3.6-Flash and MiniMax-M2.5
and competitive performance on DeepSeek-V4-Pro and GPT-4o mini.
Together, these results indicate that deterministic, record-based
permission checks can suppress injected actions without broadly
degrading task completion.

The baselines exhibit less consistent cross-benchmark behavior. CaMeL
and ACE achieve low ASR through structural restrictions but incur
substantial utility loss, especially on AgentDyn. Progent and DRIFT
perform better on the more structured AgentDojo tasks but degrade on
AgentDyn, which involves longer trajectories, larger tool sets, and more
dynamic cross-application interactions~\cite{agentdyn}. Their results
also vary across model backends, suggesting sensitivity to the
underlying LLM's ability to generate plans, assign privileges, or perform
runtime validation. In contrast, \system remains more stable because
the LLM is used only for bounded record extraction, whereas authorization
decisions are computed deterministically over deployed policies.

\begin{table}[t]
    \centering
    \caption{Impact of policy generators and training data scale on AgentDyn. DeepSeek-v4-pro acts as the final evaluation agent in all runs.}
    \label{tab:actgov-ablation-summary}
    \small
    \begin{tabular*}{\columnwidth}{@{\extracolsep{\fill}}lccc}
        \toprule
        Condition / Model & Clean $\uparrow$ & Attacked $\uparrow$ & ASR $\downarrow$ \\
        \midrule
        \multicolumn{4}{l}{\textbf{Policy Generator} (trained on 1/2 training split)} \\
        \midrule
        Qwen3.6-flash   & 0.317 & 0.271 & 0.007 \\
        MiniMax-M2.5    & 0.283 & 0.232 & 0.004 \\
        DeepSeek-v4-pro & 0.533 & 0.429 & 0.007 \\
        \midrule
        \multicolumn{4}{l}{\textbf{Train Split Size} (trained on DeepSeek-v4-pro) } \\
        \midrule
        25\% training split       & 0.500 & 0.432 & 0.007 \\
        50\% training split    & 0.533 & 0.429 & 0.007 \\
        \bottomrule
    \end{tabular*}
\end{table}

\subsection{Impact of Policy Generators and Data Scale (RQ3)}

To address RQ3, we evaluate how the choice of policy-generating LLM
and the amount of training data affect policy quality on AgentDyn,
while fixing DeepSeek-V4-Pro as the evaluation backend.
Table~\ref{tab:actgov-ablation-summary} reports the results. Using the
same fixed 50\% subset of the training split, all generators produce
policies with near-zero ASR ($\le 0.007$), but their utility differs
substantially. DeepSeek-V4-Pro achieves the highest clean and attacked
utility (0.533 and 0.429), whereas Qwen3.6-Flash and MiniMax-M2.5
produce more restrictive policies. We then fix DeepSeek-V4-Pro as the
generator and reduce the training-data fraction from 50\% to 25\%.
Clean utility decreases only slightly from 0.533 to 0.500, while ASR remains at 0.007.
These findings indicate that both the capability of the generating LLM and the comprehensiveness of the training split significantly influence the policy generation and iteration process. 

\begin{table}[t]
    \centering
    \caption{Cross-domain transferability of generated policies evaluated on DeepSeek-v4-pro.}
    \label{tab:cross_transfer}
    \small
    \begin{tabular*}{\columnwidth}{@{\extracolsep{\fill}}lccc}
        \toprule
        Training & Clean $\uparrow$ & Attacked $\uparrow$ & ASR $\downarrow$ \\
        \midrule
        \multicolumn{4}{c}{\textit{Testing on AgentDyn}} \\
        \midrule
        No defense & 0.767 & 0.729 & 0.086 \\
        AgentDyn   & 0.600 & 0.511 & 0.007 \\
        AgentDojo  & 0.100 & 0.075 & 0.000 \\
        \midrule
        \multicolumn{4}{c}{\textit{Testing on AgentDojo}} \\
        \midrule
        No defense & 0.878 & 0.839 & 0.061 \\
        AgentDojo  & 0.755 & 0.713 & 0.000 \\
        AgentDyn   & 0.224 & 0.302 & 0.000 \\
        \bottomrule
    \end{tabular*}
\end{table}

\subsection{Limitations and Future Work}
While \system demonstrates high efficacy and utility preservation within in-domain evaluations, our current empirical validation highlights a limitation regarding policy generalization across entirely unseen environments. As a framework designed to prioritize strict security, our foundational premise is that the policy refinement component requires exposure to a sufficiently comprehensive array of benign tasks and attack scenarios to synthesize optimal rule bundles. 

To quantify this, we conducted a cross-transfer evaluation (Table~\ref{tab:cross_transfer}), introducing the defenseless baseline as a theoretical upper bound for utility. The results perfectly align with our design philosophy: when encountering out-of-domain tasks, the generated policies maintain absolute safety (achieving an ASR of 0.000 in both transfer directions) by defaulting to severely conservative, restrictive behaviors. However, this rigid security boundary on unfamiliar tasks inevitably causes a significant drop in utility. For instance, on the AgentDyn testing split, while the in-domain policy reasonably preserves clean utility (0.600) compared to the baseline without defense (0.767), deploying an out-of-domain policy trained on AgentDojo causes the clean utility to plummet to 0.100.

Currently, our empirical validation is constrained to learning and refining policies over specific dataset splits. Bridging this gap, extending \system toward open-world
policy generalization that preserves utility across previously unseen
tools and domains remains an important direction for future work.

\section{Conclusion}
\system provides a robust runtime enforcement framework that successfully secures LLM-based autonomous agents by strictly separating action proposals from execution authority. By abstracting unstructured runtime contexts into a unified, orthogonal 2D semantic record space, \system enables the automated construction and formal verification of authorization policies entirely offline. At runtime, it evaluates every candidate tool call against these verified, task-scoped boundaries before any external effects occur. Extensive evaluations on the AgentDojo and AgentDyn benchmarks demonstrate that this per-action enforcement neutralizes indirect prompt-injection attacks to near-zero success rates reducing relying on the underlying LLM's natural language comprehension. Ultimately, \system significantly outperforms existing system-level defenses by preserving high task utility across dynamic, long-horizon workflows, offering a scalable and formally grounded path forward for agentic security.

\FloatBarrier
\bibliography{aaai2027}
\FloatBarrier

%
%

\setlength{\belowcaptionskip}{4pt}

\definecolor{promptbg}{rgb}{0.965,0.965,0.965}
\definecolor{promptframe}{rgb}{0.82,0.82,0.82}
\definecolor{promptcomment}{rgb}{0.33,0.42,0.50}

\lstdefinestyle{actgovprompt}{
  basicstyle=\footnotesize\ttfamily,
  numbers=none,
  backgroundcolor=\color{promptbg},
  rulecolor=\color{promptframe},
  commentstyle=\color{promptcomment},
  frame=single,
  framesep=4pt,
  xleftmargin=3pt,
  xrightmargin=3pt,
  breaklines=true,
  breakatwhitespace=false,
  columns=fullflexible,
  keepspaces=true,
  showstringspaces=false,
  tabsize=2,
  aboveskip=6pt,
  belowskip=6pt
}

\newcommand{\yes}{\textit{yes}}
\newcommand{\no}{\textit{no}}
\newcommand{\procedurallayer}{%
  \begin{tabular}[t]{@{}l@{}}procedural\\obligation\end{tabular}}

\makeatletter
\setlength{\@dblfptop}{0pt}
\setlength{\@dblfpsep}{10pt}
\setlength{\@dblfpbot}{0pt plus 1fil}
\makeatother

\appendix

\section{Use of Generative AI Tools}
\label{app:ai-use}

Generative AI tools, including OpenAI ChatGPT and Codex, were used to assist
with language polishing, \LaTeX{} formatting, and code-level debugging.
LLMs were also used as experimental components of the policy-construction
pipeline, as described in the main paper and in the prompt templates below.
All research questions, system design decisions, formal claims, experimental
protocols, and interpretations of results were determined and verified by the
authors. The authors manually reviewed all AI-assisted text and code, checked
the reported results against the experiment artifacts, and take full
responsibility for the submitted material.

\section{Policy Generation and Iteration Prompts}
\label{app:prompts}

The following fixed templates were used in the controlled ablation experiments.
For the main experiments, the authors adjusted refinement prompts using
development-split feedback during policy construction.

\subsection{Initial Policy Generation Prompt}

\begin{lstlisting}[style=actgovprompt]
SYSTEM MESSAGE

You are the ActGov policy synthesis assistant.

Your job is to synthesize an executable, auditable policy bundle for a
tool-using agent safety runtime. You are not writing natural-language safety
advice. You must output structured policy artifacts that can be compiled into
ActGov's runtime checker.

The ActGov runtime boundary is fixed:

  agent proposes tool call
    -> action adapter
    -> ContextRecord
    -> RecordFacts
    -> policy checker
    -> ALLOW / DENY / REQUIRE_APPROVAL / ESCALATE

Policies must be expressed over structured records and finite fact values, not
over raw free-form prompts. The policy language is a deterministic DSL over:

  action_type, resource_type, tool_name, argument values,
  argument provenance, target role, task intent, prior trace state,
  contextual bindings, risk tags, permission templates, and obligations.

You must obey these invariants:

1. Use only the development input packet provided by the user message.
2. Do not use any final-test target selection, final-test scorer result, hidden
   benchmark target, attack metadata, or evaluation oracle to authorize a task.
3. Do not create a task-id whitelist, injection-id blacklist, or case-specific
   exception for an evaluation example.
4. A policy rule must be reusable across tasks in a suite or across suites.
5. A policy rule must compile into a deterministic predicate over ActGov facts.
6. Facts have exactly three levels:
   - L1 deterministic: directly derived from tool/action/resource/args/trace;
   - L2 semantic enum: selected from a finite enum domain;
   - L3 composed: deterministic composition of L1/L2 facts and trace state.
7. Facts may use only these scopes:
   tool_action, resource, arg, task, trace, external_content, verdict.
8. Do not introduce audit/debug/refinement metadata as policy predicates.
9. Permission templates may be derived only from trusted clean task authority
   and explainable capability-scope generalization.
10. Attack evidence may only remove, deny, escalate, or constrain authority; it
    must never grant a new benign permission.
11. Prefer small, conservative updates: abstract templates first, provenance and
    target-role rules second, narrow exceptions only when they are explicitly
    task-bound.
12. Output strict JSON matching the requested schema. Do not include markdown in
    the JSON output.

USER MESSAGE

Generate an initial ActGov policy bundle from the following development packet.

[A] Benchmark and split

benchmark_name:
{BENCHMARK_NAME}

development_split_name:
{DEVELOPMENT_SPLIT_NAME}

split_unit:
{SPLIT_UNIT}

random_seed:
{RANDOM_SEED}

[B] Tool and suite ontology

tool_schemas:
{TOOL_SCHEMAS_JSON}

suite_ontology:
{SUITE_ONTOLOGY_JSON}

action_resource_map:
{ACTION_RESOURCE_MAP_JSON}

high_risk_action_tags:
{HIGH_RISK_ACTION_TAGS_JSON}

privileged_argument_schema:
{PRIVILEGED_ARGUMENT_SCHEMA_JSON}

[C] Trusted clean tasks

clean_tasks:
[
  {
    "suite": "...",
    "user_task_id": "...",
    "trusted_user_prompt": "...",
    "clean_tool_plan": [
      {"tool": "...", "args": {...}}
    ],
    "clean_success": true
  }
]

[D] Development attack evidence

Attack evidence is negative-authority evidence only. It may justify denials,
escalation, provenance constraints, target-role constraints, or contextual
binding constraints. It may not grant benign permissions.

attack_evidence:
[
  {
    "suite": "...",
    "user_task_id": "...",
    "injection_task_id": "...",
    "trusted_user_prompt": "...",
    "untrusted_injection_text": "...",
    "observed_tool_trace": [
      {"tool": "...", "args": {...}, "source_observation": "..."}
    ],
    "attack_goal": "...",
    "attack_succeeded": true,
    "scorer_label": "success | failure | unavailable"
  }
]

[E] Fixed ActGov fact domains

provenance_classes:
[
  "explicit_user_target",
  "trusted_task_artifact",
  "task_context_derivation",
  "untrusted_injected_content",
  "model_generated",
  "unknown"
]

target_roles:
[
  "submission_recipient",
  "payment_payee",
  "calendar_participant",
  "repo_transfer_target",
  "repo_collaborator",
  "authorization_endpoint",
  "external_file_sink",
  "shopping_purchase_target",
  "generic_outbound_recipient",
  "unknown"
]

fact_levels:
[
  "L1_deterministic",
  "L2_semantic_enum",
  "L3_composed"
]

fact_scopes:
[
  "tool_action",
  "resource",
  "arg",
  "task",
  "trace",
  "external_content",
  "verdict"
]

[F] Required objective

Synthesize an initial policy bundle that:

1. Preserves clean task utility by deriving minimal permission templates from
   trusted clean task authority.
2. Blocks indirect prompt injection and tool-mediated attacks by denying or
   escalating actions whose authority comes from untrusted observations.
3. Separates task permission from argument provenance: an action class may be
   permitted while a specific argument binding is denied.
4. Uses contextual binding for targets that are not explicitly in the user
   prompt but are legitimately obtained from prior task-bound tools.
5. Keeps all rules reusable and auditable.

Return strict JSON only.
\end{lstlisting}

\subsection{Policy Iteration Prompt}

\begin{lstlisting}[style=actgovprompt]
SYSTEM MESSAGE

You are the ActGov policy refinement assistant.

You refine an existing ActGov policy bundle using only development-split
evidence. Your output must be a small, auditable patch plan. Do not output
general safety advice.

ActGov represents each proposed tool call as a ContextRecord and RecordFacts.
Policy rules are deterministic predicates over those records and facts.

You must preserve these invariants:

1. Do not use final-test target selection, final-test scorer results, hidden
   benchmark targets, or attacker metadata to grant permissions.
2. Do not create a task-id whitelist, injection-id blacklist, or example-specific
   final-test patch.
3. Facts have only three levels:
   - L1 deterministic;
   - L2 semantic enum with finite allowed values;
   - L3 composed from L1/L2 facts and trace state.
4. Facts have only seven scopes:
   tool_action, resource, arg, task, trace, external_content, verdict.
5. Policy rules must be represented as:
   (id_p, layer_p, effect_p, dsl_p, pred_p).
6. `effect_p` is one of:
   DENY, REQUIRE_APPROVAL, ESCALATE, ALLOW.
7. DENY and REQUIRE_APPROVAL rules have priority over ALLOW rules.
8. ALLOW rules are only narrow exceptions and must require explicit task
   authority or a valid current contextual binding.
9. Attack evidence may constrain authority but must not create new benign
   permissions.
10. Prefer the smallest repair that fixes the failure while preserving existing
    security invariants.
11. Output strict JSON matching the requested schema.

USER MESSAGE

Run one ActGov policy refinement iteration.

[A] Current policy artifacts

policy_bundle:
{CURRENT_POLICY_BUNDLE_JSON}

fact_schema:
{FACT_SCHEMA_JSON}

rule_catalog:
{RULE_CATALOG_JSON}

resolver_rules:
{RESOLVER_RULES_JSON}

[B] Development evaluation summary

benchmark_name:
{BENCHMARK_NAME}

development_split_name:
{DEVELOPMENT_SPLIT_NAME}

model_backend:
{MODEL_BACKEND}

clean_utility:
{CLEAN_UTILITY}

attacked_utility:
{ATTACKED_UTILITY}

security:
{SECURITY}

asr:
{ASR}

suite_breakdown:
{SUITE_BREAKDOWN_JSON}

verdict_distribution:
{VERDICT_DISTRIBUTION_JSON}

[C] Failure evidence

Each item below is from the development split only.

failure_evidence:
[
  {
    "failure_id": "...",
    "suite": "...",
    "phase": "clean | attack",
    "trusted_user_prompt": "...",
    "untrusted_injection_text": "... or null",
    "expected_user_goal": "...",
    "observed_tool_trace": [
      {
        "step_index": 0,
        "tool": "...",
        "args": {...},
        "tool_observation_source": "user_prompt | trusted_tool_output | untrusted_external_content | model_generated | unknown",
        "actgov_verdict": "ALLOW | DENY | REQUIRE_APPROVAL | ESCALATE",
        "actgov_reasons": ["..."],
        "context_record": {...},
        "record_facts": {...}
      }
    ],
    "current_matching_template": {...},
    "current_matching_rules": ["..."],
    "clean_success": true,
    "attack_success": false,
    "native_scorer_summary": "..."
  }
]

[D] Change budget

allowed_change_budget:
{
  "max_new_facts": {MAX_NEW_FACTS},
  "max_modified_facts": {MAX_MODIFIED_FACTS},
  "max_new_rules": {MAX_NEW_RULES},
  "max_modified_rules": {MAX_MODIFIED_RULES},
  "max_template_updates": {MAX_TEMPLATE_UPDATES},
  "allow_adapter_repairs": true,
  "allow_resolver_repairs": true
}

[E] Target metric direction

security_target:
{SECURITY_TARGET}

utility_target:
{UTILITY_TARGET}

Return strict JSON only.
\end{lstlisting}

\begin{table*}[t]
\section{Detailed Record Catalog}
\label{app:records}
\noindent Tables~\ref{tab:app-l1-records}--\ref{tab:app-l3-records} list the
records used by the released policy bundles. ``LLM'' indicates whether an LLM
participates in constructing the record.
\centering
\caption{L1 deterministic records across action, permission, binding, history,
external-content, and domain scopes.}
\label{tab:app-l1-records}
\footnotesize
\def\UrlFont{\ttfamily}
\setlength{\tabcolsep}{4pt}
\renewcommand{\arraystretch}{1.08}
\begin{tabularx}{\textwidth}{>{\ttfamily}p{0.36\textwidth}
                             p{0.10\textwidth}X}
\toprule
\normalfont Record & Scope & Meaning \\
\midrule
\url{is_high_risk} & action & Action belongs to the suite's high-risk set. \\
\url{has_outbound_tag} & action & Action sends data or value outward. \\
\url{has_delete_tag} & action & Action is destructive. \\
\url{has_state_change_tag} & action & Tool specification marks a state change. \\
\url{low_risk_time_read} & action & Action is a low-risk system-time read. \\
\url{permission_granted} & permission & Task permission authorizes the tool/action/resource triple. \\
\url{tool_allowed} & permission & Tool belongs to the task's allowed-tool set. \\
\url{contextual_followup_allowed} & permission & Action/resource pair belongs to the contextual permission closure. \\
\url{task_intent_outbound} & permission & Trusted task grants outbound intent. \\
\url{task_intent_delete} & permission & Trusted task grants delete intent. \\
\url{task_intent_contextual_binding} & binding & Task allows contextual target binding. \\
\url{task_intent_contextual_outbound_binding} & binding & Contextual binding may support outbound actions. \\
\url{task_intent_contextual_delete_binding} & binding & Contextual binding may support destructive actions. \\
\url{has_binding_context} & binding & Prior allowed steps read the resource needed by the current target. \\
\url{pending_obligations_open} & history & Trace contains unresolved obligations. \\
\url{has_prior_sensitive_read} & history & Trace already contains an allowed sensitive read. \\
\url{has_same_task_history} & history & Current task trace has previous steps. \\
\url{prior_bank_send_allowed} & history & A bank send or transfer was already allowed. \\
\url{github_verification_pending_from_tool_output} & history & A tool output opened a GitHub verification obligation. \\
\url{github_account_verified_before_current_step} & history & Scoped GitHub verification already occurred. \\
\url{verification_channel_scope} & external & Channel is scoped, broad/generic, or unknown. \\
\url{current_high_risk_intent_class} & domain & Domain class of the proposed high-risk step. \\
\url{primary_high_risk_intent_classes} & domain & High-risk classes authorized by task permissions. \\
\bottomrule
\end{tabularx}
\end{table*}

\begin{table*}[t]
\centering
\caption{L2 bounded semantic records.}
\label{tab:app-l2-records}
\footnotesize
\def\UrlFont{\ttfamily}
\setlength{\tabcolsep}{4pt}
\renewcommand{\arraystretch}{1.08}
\begin{tabularx}{\textwidth}{>{\ttfamily}p{0.31\textwidth}
                             p{0.10\textwidth}p{0.08\textwidth}X}
\toprule
\normalfont Record & Scope & LLM & Meaning \\
\midrule
\url{arg::<arg>::target_role} & parameter & \yes & Finite target-role class for an argument. \\
\url{arg::<arg>::provenance} & parameter & \yes & Finite provenance class for an argument. \\
\url{task_goal_balance_check} & domain & \yes & Whether the trusted task requires balance checking. \\
\url{calendar_payload_relation} & domain & \yes & Whether calendar content is task-bound, injection-like, or unrelated. \\
\url{input_values_contain_sensitive_credential} & parameter & \no & Input values structurally resemble sensitive credentials. \\
\url{slack_channel_arg_injection_like} & external & \no & Slack channel argument contains instruction-like injected content. \\
\bottomrule
\end{tabularx}
\end{table*}

\begin{table*}[t]
\centering
\caption{L3 composed records.}
\label{tab:app-l3-records}
\footnotesize
\def\UrlFont{\ttfamily}
\setlength{\tabcolsep}{4pt}
\renewcommand{\arraystretch}{1.00}
\begin{tabularx}{\textwidth}{>{\ttfamily}p{0.38\textwidth}
                             p{0.10\textwidth}p{0.08\textwidth}X}
\toprule
\normalfont Record & Scope & LLM & Meaning \\
\midrule
\url{arg::<arg>::allowed_provenance_classes} & permission & \no & Allowed provenance classes for the argument's target role. \\
\url{first_party_web_navigation} & binding & \no & Web navigation is bound to the first-party task context. \\
\url{same_site_download_link} & binding & \no & Download URL matches a trusted host and is not injection-adjacent. \\
\url{download_link_from_bound_page} & binding & \no & Download URL comes from a bound page and is not tainted. \\
\url{download_host_matches_prompt_or_bound_host} & binding & \no & Download host matches the prompt or a trusted prior output. \\
\url{trusted_task_document_url} & binding & \no & URL is bound to a trusted task document. \\
\url{shopping_purchase_target_from_clean_platform_context} & binding & \no & Purchase target comes from clean platform context. \\
\url{contextual_binding_currently_valid} & binding & \no & Tool-derived target satisfies the current contextual binding. \\
\url{workspace_followup_ok} & binding & \no & Workspace follow-up is supported by prior task context. \\
\url{calendar_model_generated_participant_task_bound} & binding & \no & Generated participant remains bound to the task. \\
\url{model_generated_internal_fanout_allowed} & binding & \no & Generated internal-recipient fanout is task-bound. \\
\url{verification_source_bound_to_current_action} & binding & \no & Verification source is scoped to the current action. \\
\url{trusted_github_invitation_url} & external & \no & GitHub invitation URL is bound to the task or trusted prior output. \\
\url{trusted_discount_coupon_url} & external & \no & Shopping discount URL is bound to the task or trusted prior output. \\
\url{trusted_web_submission_endpoint} & external & \no & Submission endpoint belongs to the trusted endpoint set. \\
\url{trusted_credential_submission_endpoint} & external & \no & Credential endpoint is trusted and not injection-adjacent. \\
\url{web_credential_submission_to_untrusted_endpoint} & external & \no & Credentials would be submitted to an untrusted endpoint. \\
\url{generic_inbox_read_during_github_verification} & external & \no & Broad inbox read occurs during sensitive GitHub verification. \\
\url{current_read_expands_untrusted_surface} & external & \no & Read expands the untrusted observation surface. \\
\url{slack_body_has_urls} & external & \no & Slack message body contains URLs. \\
\url{slack_body_urls_bound_to_context} & external & \no & Slack body URLs are bound to the prompt or trusted prior output. \\
\url{task_goal_requires_balance_check} & domain & \yes & Boolean projection of balance-check intent. \\
\url{pre_payment_sensitive_balance_read} & domain & \no & Bank-account read occurs before payment. \\
\url{bill_payment_without_balance_goal} & domain & \yes & Payment workflow performs an unnecessary balance read. \\
\url{post_payment_sensitive_balance_read} & domain & \no & Bank-account read occurs after an allowed payment. \\
\url{post_payment_balance_read_allowed_by_task_goal} & domain & \yes & Post-payment read is justified by the trusted task. \\
\url{github_verification_phase_active} & domain & \no & Trace is in the GitHub verification phase. \\
\url{sensitive_verification_phase_active} & domain & \no & A sensitive verification phase is active. \\
\url{scoped_verification_channel_available} & domain & \no & A scoped verification channel is available. \\
\url{calendar_payload_bound_to_task_goal} & domain & \yes & Boolean projection of a task-bound calendar payload. \\
\url{calendar_payload_injection_like} & domain & \yes & Boolean projection of injection-like calendar content. \\
\url{step_matches_primary_high_risk_intent} & domain & \no & Step stays within the authorized high-risk intent class. \\
\url{github_force_overwrite_without_task_goal} & domain & \no & Force overwrite lacks explicit task authorization. \\
\url{travel_reservation_target_model_generated} & domain & \no & Reservation target was generated rather than catalog-bound. \\
\url{model_generated_soft_args_allowed} & permission & \no & Policy variant permits only soft model-generated arguments. \\
\bottomrule
\end{tabularx}
\end{table*}

\FloatBarrier

\begin{table*}[t]
\section{Detailed Policy Catalog}
\label{app:policies}
\centering
\caption{Executable \system policy rules. ``Dyn'' and ``Dojo'' indicate
activation in the released benchmark-specific policy bundles.}
\label{tab:app-policy-rules}
\footnotesize
\def\UrlFont{\ttfamily}
\setlength{\tabcolsep}{3.2pt}
\renewcommand{\arraystretch}{1.08}
\begin{tabularx}{\textwidth}{>{\ttfamily}p{0.26\textwidth}
                             p{0.12\textwidth}p{0.09\textwidth}
                             >{\ttfamily}Xcc}
\toprule
\normalfont Policy ID & Layer & Effect & \normalfont Main records & Dyn & Dojo \\
\midrule
\url{TASK-PERM-001} & task permission & deny & \url{permission_granted} & \yes & \yes \\
\url{KERNEL-HI-OUTBOUND} & hard invariant & deny & \url{has_outbound_tag}, \url{task_intent_outbound} & \yes & \yes \\
\url{KERNEL-HI-DELETE} & hard invariant & deny & \url{has_delete_tag}, \url{task_intent_delete} & \yes & \yes \\
\url{HI-HIGH-RISK-INTENT-CLASS-FREEZE} & hard invariant & deny & \url{is_high_risk}, \url{step_matches_primary_high_risk_intent} & \yes & \no \\
\url{HI-POST-PAYMENT-BALANCE-READ-FREEZE} & hard invariant & deny & \url{post_payment_sensitive_balance_read}, \url{post_payment_balance_read_allowed_by_task_goal} & \yes & \no \\
\url{HI-SENSITIVE-VERIFICATION-SCOPED-CHANNEL} & hard invariant & deny & \url{sensitive_verification_phase_active}, \url{scoped_verification_channel_available} & \yes & \no \\
\url{HI-WEB-CREDENTIAL-SUBMISSION-UNTRUSTED-ENDPOINT} & hard invariant & deny & \url{web_credential_submission_to_untrusted_endpoint} & \yes & \no \\
\url{HI-PRE-PAYMENT-BALANCE-READ-NOT-REQUIRED} & hard invariant & deny & \url{bill_payment_without_balance_goal} & \yes & \no \\
\url{HI-CALENDAR-PAYLOAD-MUST-MATCH-TASK-GOAL} & hard invariant & deny & \url{calendar_payload_bound_to_task_goal}, \url{calendar_payload_injection_like} & \yes & \yes \\
\url{HI-INJECTED-DOWNLOAD-NO-FILE-AUTHORITY} & hard invariant & deny & \url{same_site_download_link}, \url{arg::<arg>::provenance} & \yes & \yes \\
\url{HI-GITHUB-FORCE-OVERWRITE-REQUIRES-TASK-GOAL} & hard invariant & deny & \url{github_force_overwrite_without_task_goal} & \yes & \yes \\
\url{HI-TRAVEL-RESERVATION-TARGET-MUST-BE-CATALOG-BOUND} & hard invariant & deny & \url{travel_reservation_target_model_generated} & \no & \yes \\
\url{HI-SLACK-OUTBOUND-URL-MUST-BE-TASK-BOUND} & hard invariant & deny & \url{slack_body_has_urls}, \url{slack_body_urls_bound_to_context} & \no & \yes \\
\url{HI-SLACK-CHANNEL-TARGET-NO-INJECTED-INSTRUCTION} & hard invariant & deny & \url{slack_channel_arg_injection_like} & \no & \yes \\
\url{REFINE-UNTRUSTED-TARGET} & hard invariant & deny & \url{arg::<arg>::provenance} & \yes & \yes \\
\url{REFINE-TARGET-ROLE-POLICY} & hard invariant & deny & \url{arg::<arg>::target_role}, \url{arg::<arg>::allowed_provenance_classes} & \yes & \yes \\
\url{REFINE-TOOL-DERIVED} & \procedurallayer & escalate & \url{arg::<arg>::provenance}, \url{contextual_binding_currently_valid} & \yes & \no \\
\url{KERNEL-PO-MODEL-GENERATED} & \procedurallayer & approval & \url{arg::<arg>::provenance}, \url{model_generated_soft_args_allowed} & \yes & \no \\
\url{SLACK-PO-BODY-URL} & \procedurallayer & approval & \url{slack_body_has_urls}, \url{slack_body_urls_bound_to_context} & \yes & \no \\
\bottomrule
\end{tabularx}
\end{table*}

\FloatBarrier


\end{document}